\documentclass[aps,prb,amssymb,reprint,floatfix]{revtex4-1}
\usepackage{graphicx}

\usepackage{amsmath}

\begin{document}

\title{Spatial cross section of local and non-local upconversion processes for upconversion in InAs based quantum structures}
\author{David M.~Tex and Itaru Kamiya}
\affiliation{Toyota Technological Institute, Nagoya 468-8511, Japan}
\date{\today}
\begin{abstract}
Transients of upconverted photoluminescence (UPL) from molecular beam epitaxy (MBE) grown InAs quantum structures have been obtained. The results indicate the importance of spatial energy transfer cross section in the upconversion processes. The time constants determined suggest that several monolayer (ML) thick InAs quantum structures are the origin of upconversion. Both the time resolved and continuous wave data allow for the discussion of the dynamic balance between a local process (for example excitonic Auger process) and non-local process (for example two-step two-photon-absorption process with photon recycling) in such a system.
\end{abstract}
\maketitle

\section{Introduction}
\begin{figure}
\includegraphics[width=8.5cm]{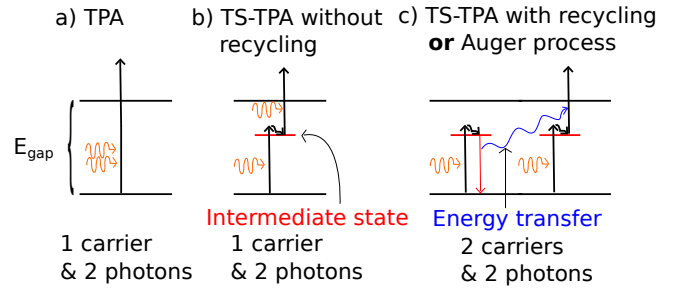}
\caption{\label{ov} (color online) Energy diagram for three types of upconversion in semiconductor systems. In all three cases two incident photons with energy about 75\% of the bandgap energy $E_{gap}$ are absorbed and the final state consists of one carrier with energy larger than $E_{gap}$. Radiative recombination from the conduction band minimum will result in an upconverted photon, that is, one photon with energy larger than the incident photons used for excitation. a) Two-photon-absorption (TPA) process. Two photons are simultaneously absorbed by a single carrier. b) Two-step two-photon-absorption (TS-TPA) without photon recycling. Successive absorption of two incident photons. The second photon must be absorbed within the time interval the carrier being in the intermediate state. Else, recombination leads to a new photon (in case of radiative recombination) with energy of the intermediate state. c) Energy transfer between two excited carriers. TS-TPA process with photon recycling is defined as the process where a photon by radiative recombination of the carrier in the left intermediate state is absorbed by the right excited carrier, {\it{i.e.}}, the dipole matrix element determines the interaction strength. In an Auger process the Coulomb potential between the carriers mediates the energy transfer.}
\end{figure}

Photon upconversion is a process in which two low energy photons are absorbed by a host material and converted into one photon of higher energy through an upconversion process. This allows for example for more efficient use of infrared light, however, the current achievable upconversion efficiencies limits its applications.\newline
\indent For application of photon upconversion it is desirable that this process takes place even in the lower power density excitation regime ($<1\ \mathrm{W/cm^2}$). Such upconversion should not occur through simultaneous absorption of two photons ((Fig.~\ref{ov}) a)), but rather through interaction of an incident photon and an excited particle (Fig.~\ref{ov} b)) or energy transfer between two excited particles (Fig.~\ref{ov} c)). Such a phenomenon is reported in semiconductor heterojunctions and bulk systems.\cite{seidel,hellmann,driessen,su,zeman,cho,cheong,johnson,quagliano,kovalev} Upconverted photoluminescence (UPL) from materials systems with InAs quantum dots (QDs) has already been reported.\cite{kammerer,paskov} Recently, we showed that upconversion is observed even for InAs thickness below QD formation and suggested the possibility of upconversion occurring through confined states in extended InAs islands.\cite{tex1} Because such quantum structures are like quantum wells (QWs) with defined borders in the in-plane directions, we term these island structures QW islands (QWIs). The observed PL excitation (PLE) spectra and power dependence suggest that in QWIs energy transfer between the excited particles (Fig.~\ref{ov} c)) is most important. Therefore, two-step two-photon-absorption (TS-TPA) processes without photon recycling (Fig.~\ref{ov} b)) will be not considered hereafter. At least two factors are involved in determining the upconversion efficiency. Firstly, the intrinsic lifetime of the intermediate state. Because interaction between two particles is needed, a longer lifetime increases the probability for energy transfer. To distinguish this time constant from the one discussed later we will call this lifetime a temporal cross section. Secondly, the spatial cross section for energy transfer between the excited particles. In the case of local energy transfer ({\it{e.g.}}~Auger processes \cite{seidel}), a large spatial overlap between the two excited particles will increase the probability for energy transfer. For non-local processes ({\it{e.g.}}~TS-TPA with photon recycling)\cite{hellmann} the cross section for the second particle for absorbing the emitted photon from the first particle has to be large. It is of interest to determine how these two cross sections contribute to upconversion in a specific system.\newline
\indent With this in mind, we would like to point out that highly efficient upconversion phenomena through rare earth\cite{auzel,suyver} are generally believed to have their origin in a long living intermediate state (high temporal cross section). In this work it is shown that high UPL efficiency can be achieved even through intermediate states with relatively short lifetimes. We explain this result as follows. Even with short-lifetime states, a highly efficient upconversion process can be observed if the spatial cross section for energy transfer between the two excited carriers is sufficiently large. With regard to the InAs based quantum structures used in this study, we believe that an efficient Auger process between excitons in InAs QWIs is the origin for UPL.\newline
\begin{figure}
\includegraphics[width=8.5cm]{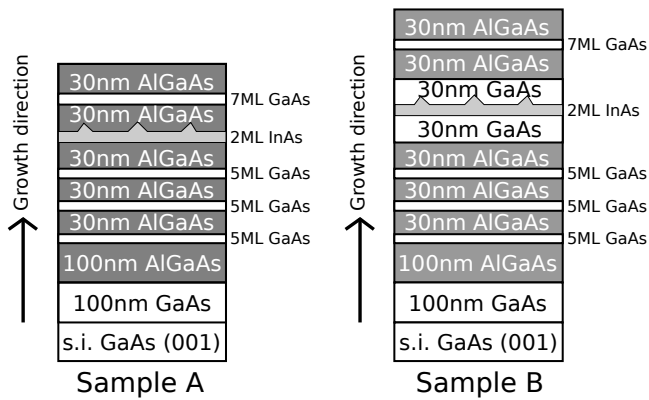}
\caption{\label{sample} Schematic description of the sample structures. In sample A the InAs layer is embedded in an AlGaAs barrier, leading to higher confined states of the InAs quantum structures compared to a GaAs barrier (sample B). The three GaAs/AlGaAs QWs below the InAs layer are referred to as the MQW, the single GaAs/AlGaAs on top is referred to as the SQW.}
\end{figure}
\section{Experiment}
In this letter, optical data obtained from two samples whose structures are shown in Fig.~\ref{sample} are compared. Detailed PL data for sample A as well as the model for the upconversion process are described in our previous publication.\cite{tex1} The intermediate states investigated here are formed in the InAs quantum structures which are sandwiched by an AlGaAs barrier (sample A) or by a GaAs barrier (sample B). Reference quantum wells (QWs) are also included in these samples.\newline
\indent Both samples were grown on semi-insulating GaAs (001) wafers by molecular beam epitaxy (MBE). Following a GaAs buffer and an Al$_{0.22}$Ga$_{0.78}$As layer growth, GaAs/AlGaAs multi QWs (MQWs) were grown at $650\ \mathrm{^\circ C}$. For sample A the growth was then interrupted and a two monolayer (ML) InAs layer was deposited at $500\ \mathrm{^\circ C}$ and immediately capped by a GaAs/AlGaAs single QW (SQW) structure while gradually raising the substrate temperature to $650\ \mathrm{^\circ C}$. For sample B two further growth interrupts were introduced; one before the InAs growth in order to grow the lower GaAs barrier at $600\ \mathrm{^\circ C}$ and another before the top SQW growth to grow the AlGaAs barrier at $650\ \mathrm{^\circ C}$.\newline
\indent Photoluminescence (PL) measurements were performed at 4K. We used a Ti:sapphire laser as a variable excitation source for Stokes and anti-Stokes PL, and PLE spectroscopy. The $532\ \mathrm{nm}$ laser line from the pump laser was supressed by a filter in front of the sample. A semiconductor laser with emission at $532\ \mathrm{nm}$ was also used for Stokes PL. Continuous wave (cw) measurements were made using a monochromator with $300\ \mathrm{mm}$ focal length and a Si-CCD detector, cooled down to $-70\ \mathrm{^\circ C}$. For time resolved PL (TRPL) the laser was mode locked for pulse operation (temporal width $7\ \mathrm{ps}$, repetition rate $80\ \mathrm{MHz}$). The signals were analyzed with a Si detector streak camera (Hamamatsu C4334).
\section{Results}
\subsection{Continuous wave measurements}
\begin{figure}
\includegraphics[width=9.5cm,clip=true,trim= 0cm 0cm 3.8cm 3.5cm]{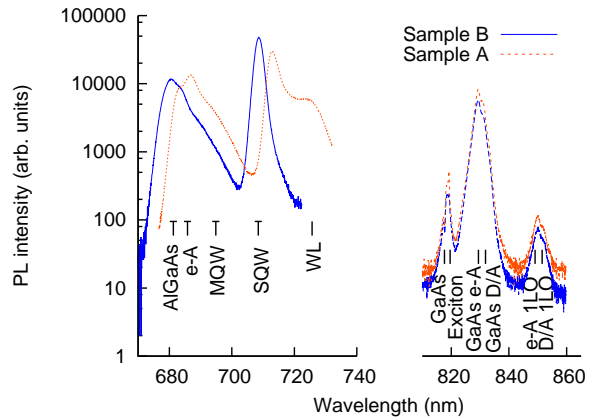}
\caption{\label{532nm} (color online) Stokes PL at $4\ \mathrm{K}$. Excitation: $532\ \mathrm{nm}$, cw $500\ \mathrm{mW/cm^2}$. The intensities are normalized by the integration time and power density. The right side shows the GaAs bandgap region, the left side the AlGaAs bandgap and the QWs.}
\end{figure}
\indent Illumination with the 532 nm cw laser ($500\ \mathrm{mW/cm^2}$) results in spectra shown in Fig.~\ref{532nm}. On the high energy side the AlGaAs related peaks and those of the QWs are found. Those of AlGaAs bandgap (AlGaAs), electron to acceptor (e-A) and QWs are indicated. On the low energy side, peaks originating from the GaAs region are observed. We assigned the peaks to the GaAs bandgap (GaAs), GaAs exciton (Exciton), GaAs electron to acceptor (GaAs e-A) and GaAs donor to acceptor (GaAs D/A) with their corresponding first longitudinal optical phonon (1LO) replica. The PL spectra from the GaAs region show similar features for both samples. The Al content for sample B was slightly higher (due to a deviation in the growth conditions), therefore, the AlGaAs band to acceptor (e-A) peak and the SQW and MQW peaks are blue shifted by $5\ \mathrm{nm}$. The InAs wetting layer (WL) peak for sample A is in the visible but not for sample B because the barrier in A is GaAs instead of AlGaAs in B, shifting the peak to longer wavelengths.\newline
\begin{figure}
\includegraphics[width=9.5cm,clip=true,trim= 0cm 0cm 3.8cm 3.5cm]{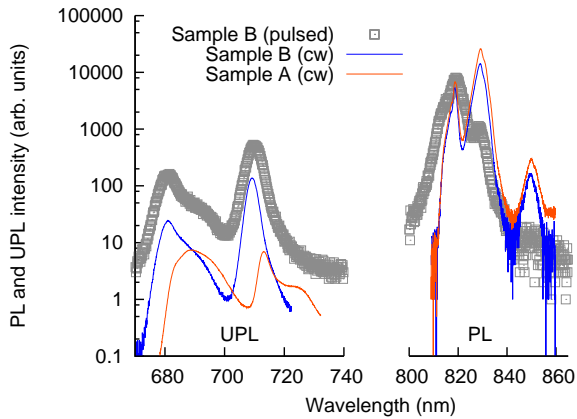}
\caption{\label{775nm} (color online) Right side: Stokes PL, left side: anti-Stokes PL (UPL). Cw-excitation: $805\ \mathrm{nm}$, $10\ \mathrm{W/cm^2}$ (anti-Stokes) and $100\ \mathrm{mW/cm^2}$ (Stokes) at 4K. The intensities are normalized by the integration time and power density. Pulsed excitation: $775\ \mathrm{nm}$, repetition rate $80\ \mathrm{MHz}$, half width $7\ \mathrm{ps}$, time averaged power density $10\ \mathrm{W/cm^2}$, normalized to the cw GaAs exciton peak. The orange and blue spectra are those obtained by cw excitation, identical to those shown in Fig.~\ref{532nm}. While the PL spectra of the QWs are almost comparable, the UPL spectra of the QWs are much stronger for sample B. The noise level of streak camera data shows the critical UPL intensity needed in order to observe time transients.}
\end{figure}
\begin{figure}
\includegraphics[width=7.5cm]{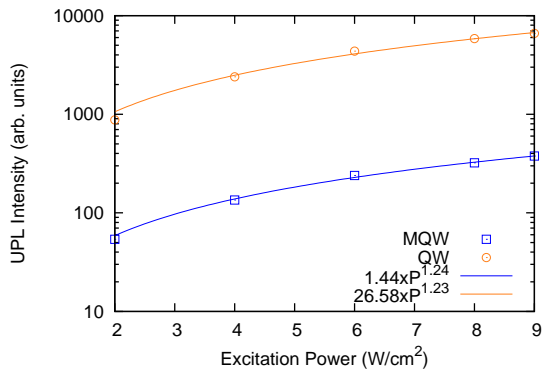}
\caption{\label{pow} (color online) Power dependence of the UPL intensities from the MQW and SQW of sample B at 4 K for excitation with $855\ \mathrm{nm}$.}
\end{figure}
\indent In order to compare the upconversion efficiencies PL measurements were performed at excitation wavelength $805\ \mathrm{nm}$ ($1.54\ \mathrm{eV}$ at $4\ \mathrm{K}$). This energy is smaller than the QW states ($>1.70\ \mathrm{eV}$), but still larger than the GaAs bandgap ($1.52\ \mathrm{eV}$). Therefore, the PL observed from the AlGaAs and QW regions are anti-Stokes PL or UPL, whereas the PL from the GaAs is Stokes PL. The data obtained by cw excitation are shown in Fig.~\ref{775nm} with the orange (sample A) and blue (sample B) curves. Again, the PL spectra from GaAs for both samples are similar. The main differences lie in the UPL. On the whole, the UPL from sample B is stronger than that from sample A. For instance, the SQW UPL from sample B is an order of magnitude larger than that from sample A. This increase in the QW UPL intensity pushes the relative UPL efficiency $\eta_{rel}$ (ratio of the SQW luminescence intensities for excitation with $1.54\ \mathrm{eV}$, $10\ \mathrm{W/cm^2}$ and $2.33\ \mathrm{eV}$, $10\mathrm{W/cm^2}$) up to about 0.2\%. The power dependence of the UPL is $\approx$1{.}23 for cw excitation (see Fig.~\ref{pow}). The gray data points in Fig.~\ref{775nm} represent the spectrum observed by the streak camera integrated over $12.5\ \mathrm{ns}$. This spectrum is normalized to the GaAs exciton peak of sample A. Due to the higher power density we see broadening of all peaks accompanied with a relative increase in the UPL intensity, which is inevitable for the streak camera measurement. \newline
\begin{figure}
\includegraphics[width=8.5cm]{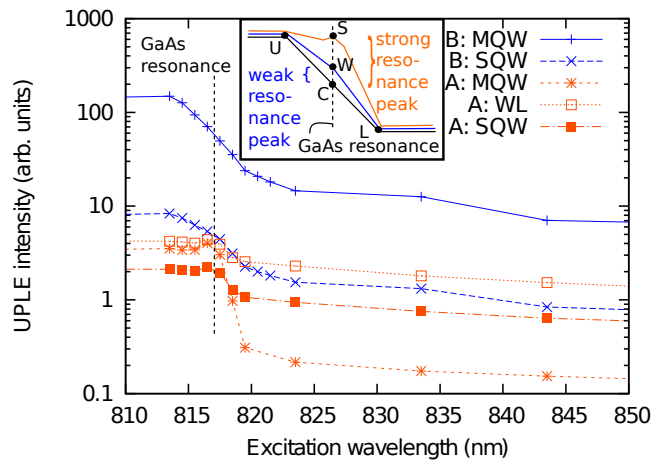}
\caption{\label{ple} (color online) PLE spectra for selected states.  Excitation power density: $10\ \mathrm{W/cm^2}$ (cw). The intensities are normalized by the integration time and power density. The data shows a clear difference between samples A and B when crossing the GaAs bandgap. From the difference in the total intensity an approximate value for the contribution of the GaAs bandgap UPL to the observed TRPL can be calculated. The lines are guides to the eyes. The inset shows how a weak and strong resonance of the GaAs bandgap UPL would be observed in the UPLE.}
\end{figure}
\indent Following our previous report,\cite{tex1} we wish to assure that the observed UPL is mainly due to the InAs layer, and not due to upconversion from the GaAs bandgap (in the following referred to as GaAs bandgap UPL) which is also excited during the TRPL measurements. This can be verified with the PLE spectra shown in Fig.~\ref{ple}, which compares both samples A and B for excitation above and below the GaAs bandgap. The features in sample A have been discussed in detail in our previous work.\cite{tex1} In brief, for sample A we observe a resonant peak at the GaAs band gap for all three states (this manifests in the plateau above the GaAs bandgap energy and a small peak at the bandgap, triangle U-S-C in Fig.~\ref{ple}), and a prominent decrease in the MQW when crossing the band gap. This is due to the capture of upconverted carriers from the GaAs buffer in the MQW states for excitation energies above the GaAs bandgap. The top SQW is less influenced, mainly receiving carriers from the InAs intermediate state. For sample B, we observe a prominent decrease for both the SQW and MQW, and only a very weak resonance at the GaAs bandgap (no plateau visible, triangle U-W-C in Fig.~\ref{ple}). This is explained by considering the nature of the resonant absorption of the GaAs barrier surrounding the InAs QWI. For wavelengths shorter than $814\ \mathrm{nm}$ GaAs is excited mainly non-resonantly, which results in part of the carriers relaxing to the InAs QWI. When the laser energy approaches the GaAs bandgap energy, stronger resonant excitation of GaAs leads to less carriers relaxing to the InAs QWI. Hence, UPL induced by the InAs layer must decrease during the resonant excitation of GaAs in sample B, which is the result observed. From the comparison between the PLE intensity increase due to the GaAs bandgap UPL (Fig.~\ref{ple}, intensity difference between points C-S and C-W for strong and weak influence, respectively) and the total intensity increase for below and above bandgap excitation (intensity difference between points L-U) we infer that for sample B the GaAs bandgap UPL contributes about eight times less (ratio of UPL through InAs QWIs and UPL through GaAs bandgap $\alpha \approx 8$) to the total UPL than the InAs QWI. This means that UPL through the intermediate state in the InAs layer should be either dominant at the initial stage of TRPL decay in sample B or has a much slower decay than the GaAs bandgap UPL. The latter case can be discarded due to the experimental data shown below.
\subsection{Time resolved measurement}
\begin{figure}
\includegraphics[width=8.5cm]{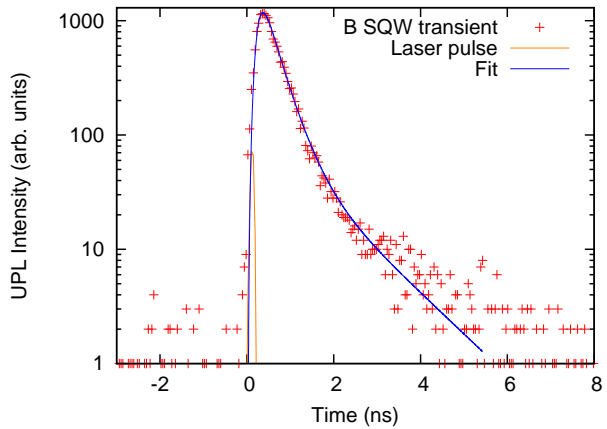}
\caption{\label{trpl} (color online) TRPL transient of the GaAs/AlGaAs SQW for sample B. Excitation: $775\ \mathrm{nm}$, repetition rate $80\ \mathrm{MHz}$, half width $7\ \mathrm{ps}$, time averaged power density $10\ \mathrm{W/cm^2}$. The intensity reaches its maximum about $270\ \mathrm{ps}$ after the laser pulse (green line). Then a clear two component decay is observed. We attribute the fast decay to the InAs UPL and the slow decay to the GaAs bandgap UPL. The fit indicated incorporates a local process (major contribution) and a non-local process (minor contribution) as described in the text. The time constants are $\tau_i=$400 ps, $\tau_{non}=$231.92 ns, $\tau_{loc}=$5.01 ns, $\tau_{d}=$1.25 ns, $\tau_{A}=$37 ns, $\tau_{relax}=$116 ns, and $\tau_{QW}=$48 ns. This curve represents 7\% upconversion efficiency (for QWI radius $r=$40 nm) with $x=$0.02 , $\alpha=$5.85 and a residual of 274{.}3, where $x$ is the ratio of non-local and local upconversion efficiency for cw excitation with $10\ \mathrm{W/cm^2}$, $\alpha$ is the ratio of UPL through InAs and GaAs, and residual meaning the deviation of the fit and the measurement as discussed later in the text.}
\end{figure}
\indent The time decay of the UPL from the top SQW of sample B is shown in Fig.~\ref{trpl}. It is obvious that the time constant for the state responsible for UPL is only on the nano-second (ns) order. To our best knowledge this is the first direct observation of the time constant for upconversion through InAs intermediate states. This data proves that upconversion with efficiencies on the order of 1\% in a semiconductor system can be realized without long lifetimes on the order of $\mathrm{\mu{}s}$, which is usually considered necessary. We explain the observed efficient UPL with short lifetime intermediate states by considering a high spatial cross section for the energy transfer between two excited particles.\newline
\indent A simple interpretation of the data is given first. The rise time of the UPL signal is about 270 ps, mainly determined by the radiative recombination time constant of the probe SQW and the exciton formation. After a maximum is reached, a two-component decay is observed. We attribute the short lifetime component (high intensity regime) to the upconversion from the InAs intermediate state and the long lifetime component (low intensity regime) to the upconversion from the GaAs bandgap region.\newline
\indent A set of rate equations is solved to determine the approximate time constants. It is assumed that a sub 20 ps process is responsible for the carrier thermalization responsible for the initial carrier redistribution.\cite{damen} The exciton gas is formed rapidly. There are two major contributions to UPL: the upconversion from the GaAs bandgap to the GaAs/AlGaAs SQW and the upconversion from the InAs quantum structure to GaAs/AlGaAs SQW. The effective decay constants for these two processes are each composed of two sub decay constants, that is,
\begin{equation}\label{basic}
\tau_e^{-1}=\tau_i^{-1}+\tau_p(n)^{-1}\ .
\end{equation}
The effective decay constant $\tau_e$ for the intermediate state for upconversion has two "decay" channels: an intrinsic lifetime $\tau_i$ (independent of the carrier density $n$), and an upconverion time constant $\tau_p$ that depends on the carrier density $n$. The effective time constants obtained from TRPL for InAs UPL and GaAs bandgap UPL are $\tau_{InAsUPL}\approx 265\ \mathrm{ps}$ and $\tau_{GaAsUPL}\approx 1300\ \mathrm{ps}$, respectively. The effective time constant for the initial increase is $\tau_{initial}\approx180\ \mathrm{ps}$. This value is a series sum of exciton formation and SQW decay. The values indicate that the time constant for the GaAs/AlGaAs SQW being within the values reported for similar systems\cite{roussignol,colocci} for an exciton relaxation time constant $\approx 100\ \mathrm{ps}$. \newline
\indent For the discussion and detailed fit of the upconversion time constants, it must be considered that these are effective values as described above. If one of the two sub time constants is much larger than the other one, the effective time constant is practically identical to the shorter sub time constant. This may be the case for the GaAs UPL. The effective time constant can be well explained with the intrinsic lifetime of carriers in the GaAs intermediate state for upconversion ($\approx 1350\ \mathrm{ps}$).\cite{texmbe} If this is the case, then the time constant for upconversion has to be one to two orders of magnitude larger, which we believe is plausible for an excitonic Auger process in bulk GaAs.\newline
\indent The focus of the following discussion lies on the InAs UPL. The effective time constant for the InAs UPL provides an important clue for elucidating the upconversion mechanisms. Time constants of Stokes PL of QDs are usually several times larger than $265\ \mathrm{ps}$, the effective time constant obtained for the InAs intermediate state.\cite{bardot,lee} If we assume that the effective time constant of the confined state in flat QDs (aspect ratio larger than $\approx$10) solely depends on the QD height,\cite{ricketts} then, considering calculations of the oscillator strength of a narrow QW,\cite{keldysh} it can be estimated that the thickness of the QW responsible for the observed InAs UPL is several times thinner than the usual QD height. This leads to the model that several monolayer thick InAs QWIs are reasonable origins for the InAs intermediate states.\newline
\indent This proposed model is supported by the following observations. InAs deposition leads to random formation of several-monolayer-thick islands before transition into QDs. Due to the stochastic growth process, islands can also exist for InAs deposition above the critical thickness for QD formation. Extended islands form thin InAs QWIs in the barrier, where at least two excitons can occupy the lowest energy state within the same island. Based on the temperature dependence and PLE of the UPL, which are similar to our earlier report, \cite{tex1} we believe that a highly efficient energy transfer between these nearly degenerated excitons confined in these thin (about 2 to 3ML) InAs QWIs is taking place.
\section{Discussion}
\begin{figure}
\includegraphics[width=8.5cm]{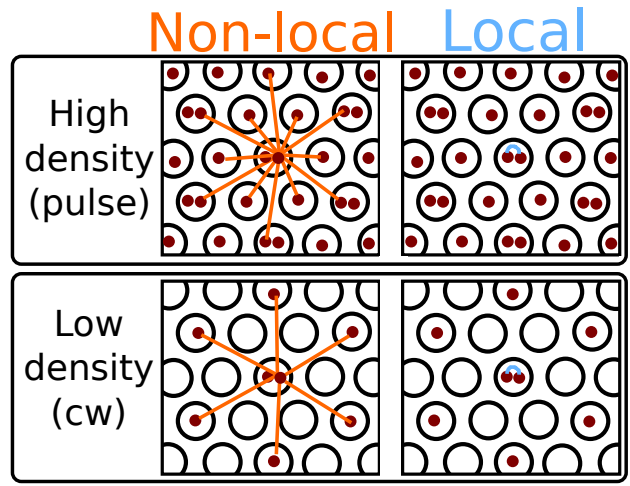}
\caption{\label{process} (color online) Schematic description of possible upconversion processes in the InAs quantum structure in top view (growth direction is out of plane). The circles and the red dots represent InAs QWIs and excitons, respectively. The lines show possible exciton-exciton interaction of the center exciton due to a non-local process (orange) or a local process (blue). In the cw case, the probability for two excitons being in the same island during the same time interval is much smaller than for the pulse excitation (both excitations have the same averaged power density), hence the probability of a second exciton being in any nearby island at any time is distinctively higher. Therefore, the non-local process must be the majority at very low excitation densities. However, if QWIs are reasonably large, the local process can be the major process down to solar excitation power densities due to higher bi-exciton formation probability (Fig.~\ref{process2}).
}
\end{figure}
\begin{figure}
\includegraphics[width=8.5cm]{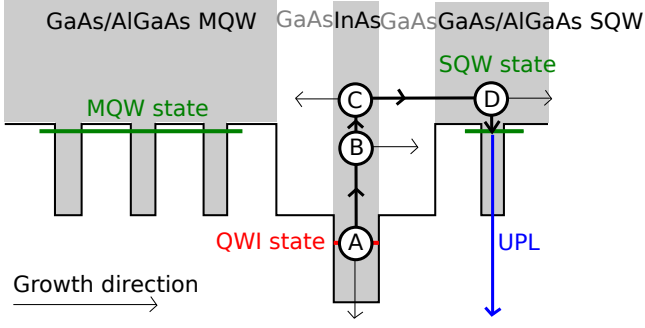}
\caption{\label{idealeff} (color online) Conduction band diagram of sample B explaining the carrier flow upon upconversion at the InAs QWI state. A) Half of the excitons taking part in the upconversion process will be upconverted. B) Due to momentum conservation, only about half of electron hole pairs receive enough energy for both electron and hole to overcome the barrier. C) Carriers will travel with 50\% probability towards the SQW. D) We assume the SQW captures these carriers with probability $75\pm25\%$. Recombination at the SQW state leads to upconverted photons (for the SQW we consider radiative recombination only).}
\end{figure}
\begin{figure}
\includegraphics[width=8.5cm]{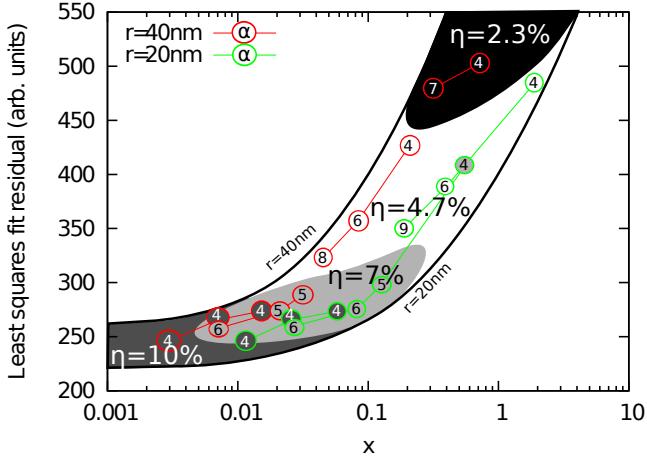}
\caption{\label{residual} (color online) The least squares fit residual is plotted as a function of $x$ (ratio between the non-local and local internal quantum efficiency for upconversion for cw excitation) for two different QWI radii (r=20 and 40 nm) and four different upconversion efficiencies ($\eta=$2.3, 4.7, 7, and 10\%). Only minima within $\alpha=8\pm4$ (the ratio of UPL through InAs and GaAs; shown in the circles) are plotted. With these points the approximate regions of equal upconversion efficiencies are determined for the investigated sample (indicated with the closed shapes in the background). Best fits, close to the experimentally obtained parameters $\alpha\approx8$ and $r\approx$20 to 40 nm are found for $\eta\approx7$\%. For $\eta\approx2.3$\% all residuals are too large, for $\eta\approx4.7$\% only residuals for large $\alpha$ are acceptable and for $\eta\approx10$\% the residuals are small, but $\alpha$ is not close enough to the estimated value. Therefore, it is suggested that the time constants obtained for $\eta\approx7$\% and $x$ on the order of 0{.}06 are most plausible (light gray region).}
\end{figure}

\indent The upconversion mechanism can be classified into two different processes: local and non-local (Fig.~\ref{process}). We assume that for both processes the ensemble upconversion rate is proportional to the average carrier density in a single island (this should be well fulfilled for low power density regime). In the non-local process, an exciton will interact with another exciton seperated up to a certain distance. For higher densities the number of partner excitons which can be found within the interaction distance is larger, therefore the recombination time decreases with increasing average carrier density. In the local process, an exciton can only interact with another exciton in the same island. It is assumed that for the excitation power densities applied, the majority of QWIs is occupied only with one or two excitons. Therefore, the local upconversion rate is expressed with a constant recombination time (equations shown below). \newline
\indent In the following we explain why we discard the non-local process from the major mechanisms in pulse excitation, and also why the contribution of both local and non-local processes may become comparable in cw excitation. \newline
\indent The definition of the upconversion efficiency is crucial for understanding the following discussion in detail. The relative intensity ratio defined earlier (ratio of the observed SQW UPL intensity for excitation at 10W/cm$^2$ and the observed Stokes PL intensity of the SQW at 10W/cm$^2$) is
\begin{equation}
\eta_{rel}=0.2{\pm0.02}\%\ .
\end{equation}
\indent With the given excitation densities the investigated structure should be
\begin{equation}
\eta_{ideal}\approx 2.4{\pm 0.8}\%
\end{equation}
in the cw case (Fig.~\ref{idealeff}). For this estimation it is assumed that the island geometry is similar to that observed from samples with 1.5ML InAs ($r_{QWI}\approx 40$ nm, area coverage $\approx 5$\%), and the absorption coefficient (for 3ML thick QWIs) is bulk-like. Here, ideal means that every exciton pair will result in an upconverted photon. However, in reality only a fraction of the excitons in the QWI contribute to upconversion. This is the internal efficiency for the upconversion mechanism,
\begin{equation}
\eta_{internal}\approx\eta_{rel}/\eta_{ideal}=8.3{\pm3.6}\%\ .
\end{equation}
This value means that out of 100 excitons in the QWIs, about 8 take part in the upconversion process, leading to 4 upconverted photons in total, that is, the internal quantum efficiency for upconversion in the QWIs observed is $\eta=4.2{\pm1.8}$\%. This efficiency may be compared with the internal quantum efficiencies of other mechanisms. The uncertainty of this value is mainly determined by the difference between the real absorption coefficient and that of bulk material, and the probability with which the SQW captures the upconverted carriers. The calculation in the following is performed for $\eta$ close to the estimated values: 2{.}3\%, 4{.}7\%, 7\%, and 10\%. It is found that only values larger than 4{.}7\% result in good fits as shown later.\newline
\begin{figure}
\includegraphics[width=8.5cm]{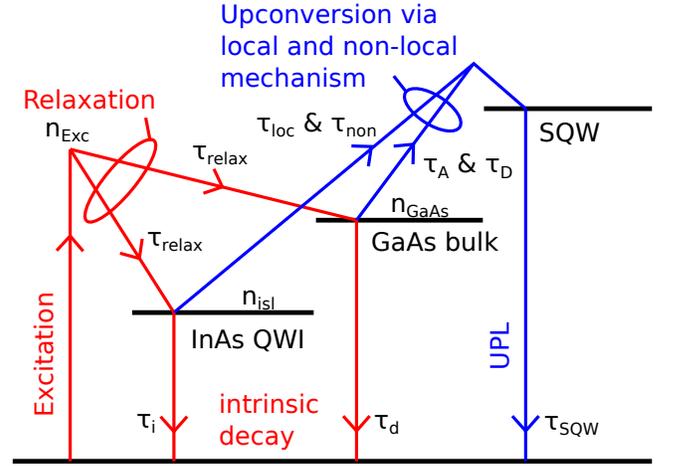}
\caption{\label{rateeq} (color online) The relaxation processes whose rate equations are discussed in the text. At the InAs and GaAs intermediate states the excited carriers can either be upconverted via a local or a non-local process, or decay radiatively or non-radiatively without transferring their energy to other excited carriers. The corresponding time constants are shown next to the arrows.}
\end{figure}
\indent The model of the relaxation processes used is shown in Fig.~\ref{rateeq}. From the left to the right it shows the excitation via the laser followed by the relaxation of the excitons to the GaAs and InAs intermediate states (for simplicity we assume the same rate for both). From each intermediate state there is a possibility to go through intrinsic decay or upconversion. The upconverted carriers are captured by the SQW, where radiative recombination results in UPL. The rate equation for the average number of excitons $n_{isl}$ in the InAs intermediate state (the QWI state) reads
\begin{equation}\label{detail}
\frac{\mathrm{d} n_{isl}}{\mathrm{d}t}=\frac{n_{Exc}}{\tau_{relax}}-\frac{n_{isl}}{\tau_{i}}-2\frac{n_{isl}}{\tau_{loc}}-2\frac{n_{isl}}{\tau_{non}}\ .
\end{equation}
This is a more specific example of the general form shown in Eq.~\ref{basic}. $n_{Exc}$ is the total number of hot excitons created by the excitation, $\tau_{relax}$ the time constant for relaxing to the ground state, and $\tau_i$ is the time constant for exciton recombination without energy transfer to another exciton. The time constants for local and non-local energy transfer between excitons are noted with $\tau_{loc}$ and $\tau_{non}$, respectively. The terms on the right side of Eq.~\ref{detail} express in order, the rate of the excitons which relax to the island, the decay of carriers due to the intrinsic lifetime, the upconversion through a local process and that through a non-local process, respectively. The equation provides the definition of the time constants used, which is needed to compare the fitted values with theoretical calculations. \newline 
\indent It is important to note that for the non-local process the time constant must depend on the density, as explained above. We define
\begin{equation}
\tau_{non}=\frac{C}{n_{isl}(t)}\ ,
\end{equation}
where C is a constant. To obtain a consistent description of the system, the fitted values have to explain the cw efficiency as well. From Eq. \ref{detail}, the equation of continuity is
\begin{equation}
\frac{\mathrm{d} n_{isl}}{\mathrm{d}t}=0=n_{abs}-n_{isl}\left(\frac{1}{\tau_i}+\frac{2}{\tau_{loc}}+\frac{2}{\tau_{non}} \right)\ ,
\end{equation}
from which the the average density can be derived. $n_{abs}$ is the number of absorbed photons per unit interval.\newline 
\indent Because this is a numerical fitting process with five parameters ($\tau_{loc}$,$\tau_{non}$,$\tau_{i}$,$\alpha$, and the intensity scaling factor), where two of them, that is, the intrinsic lifetime $\tau_{i}$ and the time constant of the local process $\tau_{loc}$ for InAs, are connected through Eq.~\ref{basic}, a unique solution cannot be determined. In order to constrain the numerous solutions with a good fit quality to a single solution we introduce a parameter describing the ratio of the upconversion contributed by the non-local process and that of the local process for cw excitation with $10\ \mathrm{W/cm^2}$:
\begin{equation}
x=\frac{\eta_{non}}{\eta_{loc}}\ .
\end{equation}
For $x=0$ the fit is for a system described only by the local process, whereas $x\rightarrow \infty$ describes a system where upconversion occurs only through a non-local process. Intermediate values are for a system where both mechanisms occur, this is the case for every real system. The condition
\begin{equation}\label{condi}
\eta_{non}+\eta_{loc}=\eta_{internal}
\end{equation}
is stringent for the validity of the fitting. In order to validate the fit, $\eta_{non}$ is varied between $0$ and $\eta_{internal}$ and $\eta_{loc}$ is calculated from Eq.~\ref{condi}. The resulting theoretical decay curve as a function of $x$ and the real data are compared. The difference is expressed with the residual squares (Fig.~\ref{residual}).
 Best fits are found for $x<0.2$ and ratios of UPL through InAs QWIs and GaAs bulk are $\alpha>4$. Therefore, the local process must be dominating in the pulse regime and also playing an important role in the cw regime. The average fitting values for the points in the 7\% region in Fig.~\ref{residual} is $x=0.048\pm0.036$, $\alpha=5.76\pm0.45$, $\tau_{i}=383{\pm22}\ \mathrm{ps}$, $\tau_{loc}=4{.}9{\pm0{.}3}\ \mathrm{ns}$ and $\tau_{non}\approx{203.7\pm129.0}\ \mathrm{ns}$.  Lower efficiencies (4.7\%) result in roughly values doubled for $x$, $\alpha$, and $\tau_{loc}$. This suggests that any reasonable time constant for the local process is smaller than 10 ns, several times smaller than the non-local time constant.\newline
\indent The estimated range of $x$ predicts a power dependence close to linear for cw excitation with $10\ \mathrm{W/cm^2}$, as is observed in the experiment.
\begin{figure}
\includegraphics[width=8.5cm]{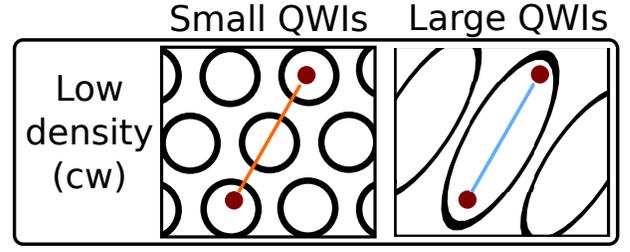}
\caption{\label{process2} (color online) Interactions between two excitons (red dots) in small an large QWIs (circles and ellipses, respectively) for low excitation power densities. In systems with small QWIs the distant excitons can only interact via non-local dipole interaction. For extended QWIs the probability for two excitons in the same QWI is higher. Therefore, local processes can be major under small excitation densities. The exciton radius and exciton diffusion determine if the local process is a dipole or an Auger process.
}
\end{figure}
\indent Such a system ({\it{i.e.}}, a system with dynamic relation between local and non-local processes) will gradually change its power dependence from quadratic (low power density) to linear (high power density) also observed by other groups.\cite{paskov} Samples with large island densities should then have a quadratic power dependence up to high excitation densities, whereas samples with low island densities show linear dependence even for lower excitation densities (Fig.~\ref{process2}) (we find the crossover between squared and linear power dependence at about $1\ \mathrm{W/cm^2}$). From the experimental point of view this model seems to account for all observed upconversion characteristics. We suggest two elementary mechanisms for the local and non-local processes. Excitonic Auger processes are purely local, recombination of an exciton will transfer its energy to the second exciton under momentum conservation. This exciton will then break up into an upconverted electron and hole. The linear power dependence of the Auger process is only strictly correct as long as the number of bi-excitons in QWIs is proportional to the excitation power density, and three or more excitons do not occupy the lowest energy state in the QWI. The observed power dependence 1.23 may also be interpreted such that tri-exciton formation takes place to some extent.\newline
\indent The non-local process can be explained with a TS-TPA process with photon recycling, {\it{i.e.}}~, two excitons interact resonantly via electromagnetic dipole field. This process also inherits the ability for local upconversion, however, if the time constant for an excitonic Auger process is much shorter, local upconversion essentially occurs through Auger and the non-local part is described by the TS-TPA process with photon recycling. To verify the correctness of this assignment a theoretical study on the upconversion time constant is needed. This requires the matrix elements to be determined without a large error due to the uncertainty of the material parameters.
\section{Summary}
\indent To conclude, we found that the time constant responsible for efficient UPL (experiment: $\eta_{rel}\approx$0.2\% for a state at $\approx 1.77\ \mathrm{eV}=700\ \mathrm{nm}$ with an excitation energy of $1.54\ \mathrm{eV}=805\ \mathrm{nm}$, that is, $0.2\ \mathrm{eV}$ energy gain with a power density of $10\ \mathrm{W/cm^2}$) in InAs quantum structures is on the order of 265 ps (this is the effective time constant). This is relatively short compared to the usually expected lifetimes for efficient UPL materials systems. We estimate that this system can achieve $4.2{\pm1.8}\%$ internal quantum efficiency for upconversion, which is a promising value considering the small excitation power density. The order of the estimated efficiency is confirmed with the fit of the UPL transient ($\approx 7$\%), suggesting an intrinsic lifetime of $\approx 380$ ps and an upconversion time constant of $\approx 4.9$ ns.\newline
\indent We point out the importance of considering high spatial cross section processes for energy transfer for UPL. For InAs quantum structures such as those we have grown, the spatial cross section is relatively large compared to the temporal cross section. We define the onset of relative strong spatial cross section for $\tau_{loc}<30\tau_i$. The fitted value $\tau_{loc}\approx 4.9$ ns is less than 30 times larger than $\tau_i\approx 380$ ns, which is close enough to allow for high UPL efficiency by energy transfer. The statement of strong spatial cross section holds true for any upconversion model conjectured for this system, because despite the relatively good UPL efficiency the UPL decay time constant is below 1 ns. We explain this with the high spatial overlap of the excitons in the QWIs, and low impurity concentration in the crystal. A numerical investigation of the data leads to the conclusion that non-local processes (for example TS-TPA with photon recycling) are minor in this materials system for excitation with $10\ \mathrm{W/cm^2}$. Local processes (for example excitonic Auger) can be used to explain both the cw and pulse data correctly, yet a non-local process may play a supporting role in a cw regime where bi-exciton formation becomes less frequent.\newline
\indent The total efficiency is proportional to the product of the spatial and temporal cross sections. Naturally, one needs to maximize this product to achieve UPL at high efficiency. However, not all materials systems allow significant increase in the temporal cross section (lifetime). Hence, the key factor to be investigated is the spatial cross section when developing upconversion devices with semiconductor materials. Intelligent structures could lead to much better spatial cross sections, allowing UPL efficiencies to reach a level useful for applications.

\begin{acknowledgments}
This work was supported by the Strategic Research Infrastructure
Project, the Ministry of Education, Culture, Sports, Science and Technology,
Japan.
\end{acknowledgments}

\bibliographystyle{apsrev}

\end{document}